\documentclass[11pt]{article}

\usepackage[preprint]{acl}
\usepackage{booktabs} 
\usepackage{multirow}

\usepackage{times}
\usepackage{latexsym}
\usepackage{amsmath} 
\usepackage{amssymb} 

\usepackage[T1]{fontenc}

\usepackage[utf8]{inputenc}

\usepackage{microtype}

\usepackage{inconsolata}

\usepackage{graphicx}
\usepackage{subcaption}

\title{GLIER: Generative Legal Inference and Evidence Ranking for Legal Case Retrieval}

\author{
Minghan Li\thanks{Equal contribution.}\thanks{Corresponding author.},
Tianrui Lv\footnotemark[1],
Chao Zhang,
Guodong Zhou \\
Soochow University, Suzhou, China \\
\texttt{mhli@suda.edu.cn, trlvtrlv@stu.suda.edu.cn}, \\ \texttt{czhang1@stu.suda.edu.cn, gdzhou@suda.edu.cn}
}

\begin{document}
\maketitle
\begin{abstract}
The semantic gap between colloquial user queries and professional legal documents presents a fundamental challenge in Legal Case Retrieval (LCR). Existing dense retrieval methods typically treat LCR as a black-box semantic matching process, neglecting the explicit juridical logic that underpins legal relevance. To address this, we propose GLIER (Generative Legal Inference and Evidence Ranking), a framework that reformulates retrieval as an inference process over latent legal variables. GLIER decomposes the task into two interpretability-driven stages: (1) A Joint Generative Inference module that translates raw queries into latent legal indicators (Charges and Legal Elements), employing a unified sequence-to-sequence strategy where charges and elements are generated jointly to enforce logical consistency; and (2) A Multi-View Evidence Fusion mechanism that aggregates generative confidence with structural and lexical signals for precise ranking. Extensive experiments on LeCaRD and LeCaRDv2 demonstrate that GLIER outperforms strong baselines like SAILER and KELLER. Notably, our framework exhibits exceptional data efficiency, maintaining robust performance even when trained with only 10\% of the data.
\end{abstract}

\section{Introduction}
\begin{figure}[t]
    \centering 
    \includegraphics[width=\linewidth]{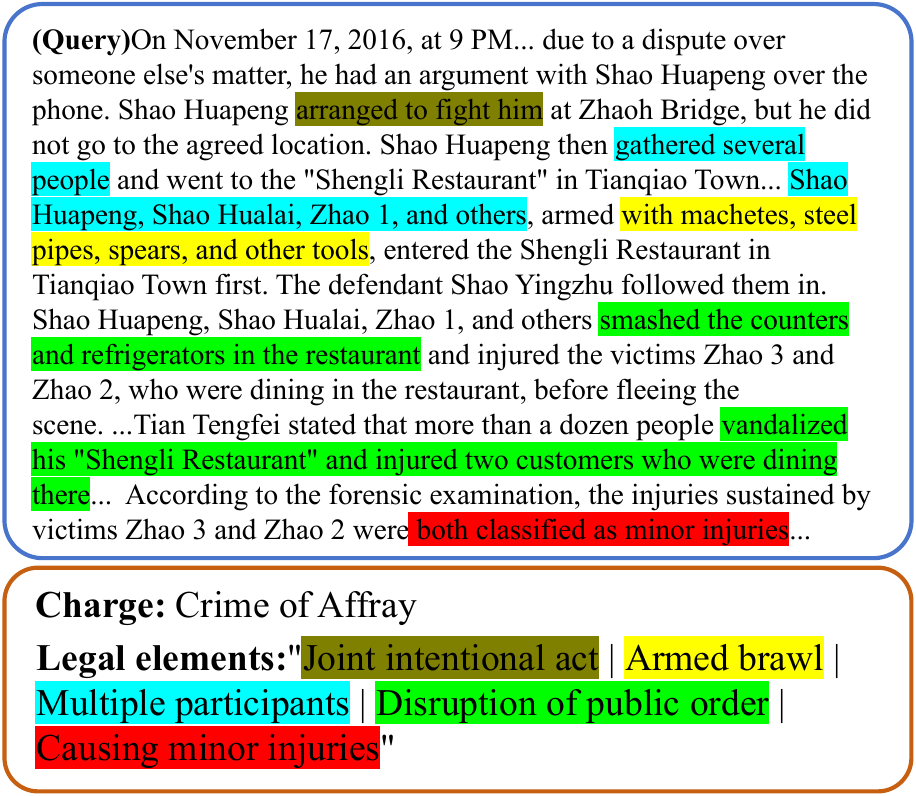} 
    \caption{A colloquial query must be mapped to structured legal concepts (e.g., charge and constitutive elements) to retrieve legally relevant precedents.}
    \label{fig:case_example} 
\end{figure}
Legal Case Retrieval (LCR) aims to identify legally relevant precedents from a large corpus given a query case \cite{feng-etal-2024-legal,t-y-s-s-hernandez-2025-lexkeyplan}. Unlike general ad-hoc retrieval, legal relevance is determined not by surface-level semantic similarity, but by whether cases share consistent juridical interpretations. In particular, relevance hinges on the alignment of \emph{Charges} and their associated \emph{Constitutive Elements}, which encode the legal logic underlying a conviction. This makes LCR challenging due to a pronounced semantic gap: queries are often colloquial factual narratives, while candidate cases are written in formal and highly structured legal language.

Existing approaches to LCR mainly follow three paradigms. Lexical matching methods such as BM25 capture explicit keywords but fail to model legal reasoning. Dense retrieval models based on pre-trained language models (PLMs) improve semantic matching, yet struggle with long documents and implicit juridical structure. More recently, Generative Retrieval (GR) \cite{li2023multiviewidentifiersenhancedgenerative,li2023learningrankgenerativeretrieval,tang2024caselinkinductivegraphlearning}methods directly generate document identifiers, but suffer from limited interpretability and hallucination risks, which are particularly problematic in high-stakes legal scenarios. A common limitation of these approaches is that they treat retrieval as a direct mapping from queries to documents, without explicitly modeling the legal reasoning process that mediates relevance \cite{deng-etal-2024-element}.

We argue that legal case retrieval should instead be formulated as inference over latent juridical structures. Legal experts typically begin by inferring legal interpretations from the facts, such as the applicable charges and their elements, and then verify these interpretations against relevant precedents. Motivated by this process, we reformulate LCR as a retrieval problem with \emph{structured latent variables}, where legal relevance is mediated by an inferred legal interpretation rather than determined by direct textual similarity.

Based on this formulation, we propose \textbf{GLIER}, a Generative Legal Inference framework for legal case retrieval. Instead of relying on complex multi-stage pipelines, GLIER infers a latent legal interpretation from the query via a \emph{unified sequence-to-sequence generation strategy}. By training the model to predict the charge and its constitutive elements as a single joint sequence, we leverage the autoregressive nature of the decoder to enforce logical consistency: the generation of legal elements is implicitly conditioned on the preceding charge prediction. The inferred latent structure is then used to mediate evidence-based ranking of candidate documents by combining generative confidence with structural and lexical matching signals, enabling interpretable and robust retrieval.

We evaluate GLIER on two benchmarks, LeCaRD \cite{10.1145/3404835.3463250} and LeCaRDv2 \cite{li2023lecardv2largescalechineselegal}. Experimental results show that GLIER consistently outperforms strong baselines such as SAILER \cite{li2023sailer} and KELLER \cite{deng2024learning} on both datasets. In particular, GLIER achieves the best overall performance on LeCaRDv2 under the same experimental setting, while substantially improving recall- and hit-oriented metrics on LeCaRD. Notably, GLIER maintains strong performance even when trained with only 10\% of the available data, demonstrating high robustness and data efficiency.

Our contributions are summarized as follows:
\begin{itemize}
    \item We formalize legal case retrieval as inference over \emph{structured latent legal variables}, explicitly modeling charges and constitutive elements as pivotal mediators of relevance.
    \item We propose a \emph{joint generative inference} framework that approximates latent legal reasoning via a unified sequence-to-sequence paradigm, guaranteeing both interpretability and logical consistency.
    \item We empirically validate that integrating latent inference with lightweight evidence-based ranking yields robust improvements. Notably, our model demonstrates exceptional data efficiency, maintaining superior performance even when trained on only 10\% of the data.
\end{itemize}

\section{Related Work}
\subsection{Legal Case Retrieval (LCR)}
Traditional Legal Case Retrieval methods mainly rely on lexical matching, such as BM25, which remains highly competitive in capturing precise keywords in legal documents. However, legal cases are often extremely lengthy and contain a large amount of specialized terminology, making it difficult for traditional methods to capture deep semantic matches. With the development of pre-trained language models (PLMs), dense retrieval-based methods have gradually become mainstream. Models such as BERT \cite{devlin-etal-2019-bert} and Lawformer \cite{xiao2021lawformer} process long documents through paragraph-level interactions or long-document attention mechanisms. SAILER \cite{li2023sailer} further introduces a structure-aware pre-training objective, enhancing representation learning by utilizing the reasoning and decision sections of cases. Despite significant progress made by these discriminative models, they typically rely on truncating or segmenting long documents, which can result in the loss of the case's global context and key legal features, such as the logical connections between charges and legal elements.

\subsection{Knowledge-Guided Case Reformulation}
To address the noise and computational redundancy caused by long documents, recent research has begun to use large language models (LLMs) to reformulate or summarize cases \cite{gao-etal-2024-enhancing-legal}. PromptCase \cite{tang2023prompt} uses LLMs to extract 'legal facts' and 'legal issues' from cases as key features, replacing the full text for encoding. Recently, KELLER \cite{deng2024learning} further proposed a knowledge-guided reformulation method that leverages LLMs to transform complex case details into concise 'crime-subfact' pairs and conducts multi-granularity contrastive learning based on these subfacts. Although these methods effectively extract core information (such as charges and legal elements) through LLMs, they essentially still fall under the discriminative retrieval (Retrieve-then-Rank) paradigm, which relies on dual encoders to compute similarity scores between the Query and the Document. This approach requires calculating similarity scores for a vast number of candidate vectors during inference and fails to directly model the generation probability from the Query to the Document's core features.

\subsection{Generative Reasoning for Legal Retrieval}
\label{sec:rel_gen_reasoning}

Recently, Generative Retrieval (GR) has emerged as a new paradigm in information retrieval, where models like DSI \cite{tay2022transformermemorydifferentiablesearch} and NCI \cite{wang2023neuralcorpusindexerdocument} directly generate document identifiers (DocIDs) to bypass the traditional index-and-retrieve pipeline. In the legal domain, models such as LegalSearchLM \cite{kim2025legalsearchlm} have explored this direction by mapping queries to case IDs. However, these methods often struggle with the "hallucination" problem and the lack of fine-grained evidence alignment, which are critical in professional legal scenarios.

Distinct from pure GR that aims at identifier generation, our work aligns with the emerging trend of \textbf{Generative Reasoning for Ranking}. This direction focuses on utilizing the zero-shot or few-shot reasoning capabilities of Large Language Models (LLMs) to expand queries or infer latent variables. Specifically, our framework treats generative models not as an end-to-end retriever, but as a \textit{semantic bridge} that translates colloquial queries into structured legal indicators (e.g., charges and constitutive elements). Unlike previous methods like PromptCase \cite{tang2023prompt} or KELLER \cite{deng2024learning} that primarily use LLMs for query reformulation, our approach explicitly models the hierarchical relationship between legal concepts and incorporates generative confidence into a discriminative fusion layer. This strategy combines the interpretability of generative inference with the robustness of traditional evidence-based ranking.

\begin{figure*}[t!]
    \centering
    \includegraphics[width=\textwidth]{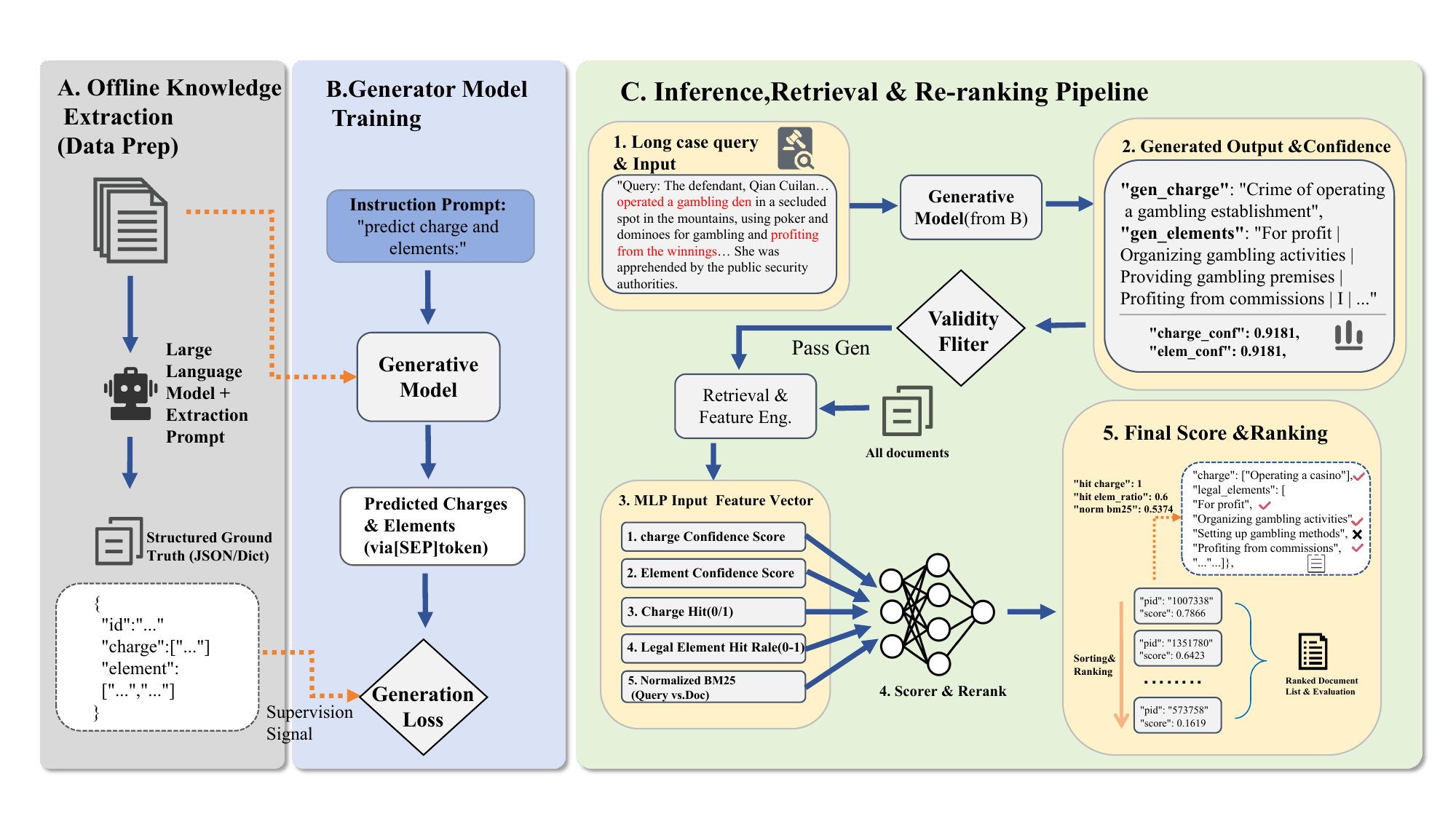}
    \caption{The overall architecture of the proposed framework, consisting of the Generative Legal Indicator Extractor (GLIE) and the Multi-Faceted Discriminative Re-ranker (MFDR).}
    \label{fig:bigpicture}
\end{figure*}

\section{Methodology}

\subsection{Problem Formulation}
Let $\mathcal{Q}$ denote the set of query cases and $\mathcal{D}$ the corpus of candidate documents. Given a query $q \in \mathcal{Q}$, Legal Case Retrieval (LCR) aims to rank documents $d \in \mathcal{D}$ by their legal relevance to $q$.

We model legal relevance as being mediated by a latent juridical structure rather than direct text similarity. Specifically, we introduce a structured latent variable $z=(c,e)$, where $c$ denotes a legal charge and $e$ denotes its associated constitutive elements. We assume that the relevance between a query $q$ and a document $d$ can be assessed through the consistency between $d$ and a plausible juridical interpretation $z$ inferred from $q$. Formally, we first infer the most probable latent structure
\begin{equation}
\hat{z} \;=\; \arg\max_{z} P_{\theta}(z \mid q),
\end{equation}
and then define the relevance score as
\begin{equation}
S(q,d) \;=\; f_{\psi}(q,d,\hat{z}),
\end{equation}
where $P_{\theta}(z \mid q)$ infers a latent legal interpretation from the query, and $f_{\psi}$ is a scoring function that aggregates multiple evidence signals conditioned on $\hat{z}$.

Legal reasoning exhibits a logical dependency, where the admissible constitutive elements are intrinsically constrained by the applicable charge. We capture this dependency by decomposing the latent distribution via the chain rule:
\begin{equation}
P_{\theta}(z \mid q) \;=\; P_{\theta}(c \mid q)\,P_{\theta}(e \mid q,c).
\end{equation}
In practice, we approximate $\hat{z}$ via \textbf{joint generative inference}: we train a sequence-to-sequence model to generate $c$ and $e$ as a unified sequence (i.e., $c \oplus [\text{SEP}] \oplus e$), which enables the model to implicitly learn the conditional dependency $P_{\theta}(e \mid q,c)$ through autoregressive decoding. We obtain $\hat{z}$ using constrained beam search and apply a validity filter based on a legal taxonomy to reduce hallucinated structures.

\subsection{LLM-driven Legal Knowledge Distillation}
\label{sec:distillation}

Legal documents often contain verbose narratives, redundant procedural details, and noise, posing significant challenges for direct dense retrieval. To mitigate this and construct a high-quality supervision signal for our student model, we employ a Large Language Model (LLM) as an offline \textbf{Knowledge Teacher} to distill structured legal signals from the corpus $\mathcal{D}$.

Specifically, we utilize \textbf{ChatGLM} (Du et al., 2022), a robust bilingual LLM, to extract core juridical components from each document $d \in \mathcal{D}$. We construct a domain-specific prompt $\mathcal{P}$ that enforces strict constraints to ensure the validity of the "Silver Standard" data:

\begin{itemize}
    \item \textbf{Terminology Enforcement:} The model is restricted to extracting \textit{Constitutive Elements} using professional legal terminology (e.g., "secretly taking property") rather than vague descriptive phrases.
    \item \textbf{Prevention of Target Leakage:} Crucially, the prompt explicitly instructs the model to exclude sentencing outcomes (e.g., "fixed-term imprisonment", "compensation") and post-crime procedural details. This ensures that the retrieved features are based solely on the \textit{facts of the crime}, preventing the model from cheating by matching sentencing patterns.
\end{itemize}

For a document $d$ with a grounded charge $c_{gt}$, the distillation process is formulated as:
\begin{equation}
    K_{d} = \text{LLM}(d, c_{gt}, \mathcal{P}) = (c_{d}, e_{d})
\end{equation}
where $c_{d} \subseteq \mathcal{K}_{charge}$ denotes the applicable charges, and $e_{d} \subseteq \mathcal{K}_{element}$ represents the extracted constitutive elements. The output is parsed from a structured JSON format. This process transforms unstructured legal texts into a clean, structured "Silver Standard" dataset $\mathcal{D}_{struct} = \{(d, c_d, e_d)\}$, providing explicit supervision for the subsequent student model without requiring expensive human annotation. The detailed prompt design is provided in ~\ref{sec:Datasets}.

\subsection{Generative Legal Inference (The Student Model)}
\label{sec:student_model}
To equip the retriever with legal reasoning capabilities, we train a sequence-to-sequence model (based on mT5~\cite{xue2021mt5massivelymultilingualpretrained}) to mimic the extraction process. Instead of independent classification, we propose a \textbf{One-Step Joint Generation} strategy to model the inherent dependencies between charges and elements.

\subsubsection{Joint Generation Training}
We formulate the task as generating the structured tuple $K_q = (c_q, e_q)$ given the query text $q$. The input $X$ is the raw query prepended with a task prompt, and the target $Y$ concatenates the charge and elements using a special separator token:
\begin{equation}
    Y = c_q \oplus \text{\texttt{[SEP]}} \oplus e_q
\end{equation}
The model is optimized by minimizing the negative log-likelihood of the target sequence:
\begin{equation}
    \mathcal{L}_{\text{gen}} = - \sum_{t=1}^{|Y|} \log P(y_t | y_{<t}, X; \theta)
\end{equation}
This joint modeling allows the decoder to leverage the predicted charge as a condition for generating subsequent legal elements, effectively preserving the logical consistency of legal reasoning.

\subsubsection{Constraint-Aware Inference}
During inference, given a query $q$, the model generates a raw sequence $\hat{Y}_q$. To mitigate hallucination inherent in generative models, we apply a \textbf{Validity Constraint Mechanism}. The raw output is parsed into candidate terms $(\hat{c}_{\text{raw}}, \hat{e}_{\text{raw}})$ and filtered against the predefined taxonomy $\mathcal{K}$:

\begin{equation}
\begin{split}
    \hat{c}_q &= \{ t \in \hat{c}_{\text{raw}} \mid t \in \mathcal{K}_{charge} \}, \\
    \hat{e}_q &= \{ t \in \hat{e}_{\text{raw}} \mid t \in \mathcal{K}_{element} \}
\end{split}
\label{eq:constraint}
\end{equation}
This ensures that the inferred knowledge is legally valid while retaining the model's high-confidence predictions.

\subsection{Multi-View Evidence Fusion Mechanism}
\label{sec:fusion}
While the generative model captures semantic reasoning, it lacks the calibration for fine-grained ranking. We propose a lightweight \textbf{Multi-View Scorer} that fuses signals from three perspectives: \textit{Latent Confidence}, \textit{Explicit Structure}, and \textit{Lexical Matching}.
For a query-document pair $(q, d)$, we construct a feature vector $\mathbf{v}_{q,d} \in \mathbb{R}^5$:

\paragraph{1. Latent Confidence View ($v_1, v_2$):}
These features quantify the generator's internal certainty regarding the inferred legal concepts. We compute the length-normalized probability for the generated charge and element sequences:
\begin{equation}
\begin{split}
    v_1 &= \exp\left( \frac{1}{|\hat{c}_q|} \sum_{t} \log P(t | \hat{c}_{<t}, q) \right), 
\\
    v_2 &= \exp\left( \frac{1}{|\hat{e}_q|} \sum_{t} \log P(t | \hat{e}_{<t}, q) \right)
\end{split}
\label{eq:v1v2}
\end{equation}
A higher probability ($v_1, v_2 \approx 1$) indicates the model has correctly identified robust legal patterns in the query.

\paragraph{2. Explicit Structural View ($v_3, v_4$):}
This view measures the overlap between the query's inferred knowledge $(\hat{c}_q, \hat{e}_q)$ and the document's ground truth $(c_d, e_d)$:
\begin{equation}
    v_3 = \mathbb{I}(\hat{c}_q \cap c_d \neq \emptyset), \quad v_4 = \frac{|\hat{e}_q \cap e_d|}{|\hat{e}_q| + \epsilon}
\end{equation}
Here, $v_3$ is a binary indicator of charge matching (a prerequisite for legal relevance), and $v_4$ represents the element support ratio.

\paragraph{3. Lexical Matching View ($v_5$):}
To incorporate traditional keyword signals, we use BM25. Crucially, to handle score variations across queries, we apply \textbf{Per-Query Normalization} using the maximum score within the candidate pool $\mathcal{C}_q$:
\begin{equation}
    v_5 = \frac{\text{BM25}(q, d)}{\max_{d' \in \mathcal{C}_q} \text{BM25}(q, d')}
\end{equation}

\subsubsection{Scoring and Optimization}
The fusion scorer is instantiated as an MLP that maps $\mathbf{v}_{q,d}$ to a relevance score $S(q, d)$. 
To improve discriminative power, we employ a \textbf{Hard Negative Mining} strategy. Instead of random sampling, we select hard negatives $\mathcal{N}_{hard}$ from top-ranked non-relevant documents retrieved by BM25. These documents share high lexical overlap with the query but differ in legal characterization.
The model is trained via Binary Cross-Entropy (BCE) loss to distinguish positive document $d^+$ from hard negatives:

{\small
\begin{equation}
    \mathcal{L}_{\text{score}} = - \left[ \log S(q, d^+) + \sum_{d^- \in \mathcal{N}_{hard}} \log (1 - S(q, d^-)) \right]
\end{equation}
}
This forces the model to look beyond keyword matching ($v_5$) and rely on structural evidence ($v_1 \dots v_4$) to distinguish subtle legal differences.

\section{Experiment}
In this section, we conduct comprehensive experiments to evaluate our proposed framework, focusing on the following research questions: 
\textbf{RQ1}: How does our framework compare against state-of-the-art baselines? 
\textbf{RQ2}: What are the contributions of the hybrid scoring mechanism and different evidentiary signals (lexical vs. generative) to the ranking performance? 
\textbf{RQ3}: How robust is the model under low-resource training settings?
\textbf{RQ4}: Does the hierarchical joint generation strategy outperform independent prediction? 

\subsection{Experimental Setup}

\subsubsection{Datasets and Evaluation Metrics}
We evaluate our method on two widely used benchmark datasets: \textbf{LeCaRD} and \textbf{LeCaRDv2} (Legal Case Retrieval Dataset).

Following standard protocols, we consider cases with a relevance label of 3 in LeCaRD and labels of 2 and 3 in LeCaRDv2 as positive. We report a comprehensive set of metrics including MAP, P@3, R@3, R@5, Hits@3, Hits@5, and MRR@5 to evaluate both ranking quality and recall capabilities.

\subsubsection{Baselines}
We compare our method with comprehensive baselines categorized into three groups: (1) \textbf{Traditional Models} including BM25 and TF-IDF; (2) \textbf{PLM-based and Embedding Methods}, encompassing general encoders (BERT, RoBERTa, BGE) and legal-specific pre-trained models (Lawformer, SAILER); and (3) \textbf{Generative/Reformulation Methods} represented by KELLER, a state-of-the-art approach utilizing LLMs for query augmentation.

All PLM-based or legal model baselines (e.g., BERT, RoBERTa, Lawformer) are fine-tuned on the respective training sets.

\begin{table*}[t!]
\centering
\resizebox{\textwidth}{!}{
\begin{tabular}{l|ccccccc|ccccccc}
\toprule
\multirow{2}{*}{\textbf{Model}} & \multicolumn{7}{c|}{\textbf{LeCaRD}} & \multicolumn{7}{c}{\textbf{LeCaRDv2}} \\
\cmidrule(lr){2-8} \cmidrule(lr){9-15}
 & \textbf{MAP} & \textbf{P@3} & \textbf{R@3} & \textbf{R@5} & \textbf{Hits@3} & \textbf{Hits@5} & \textbf{MRR@5} & \textbf{MAP} & \textbf{P@3} & \textbf{R@3} & \textbf{R@5} & \textbf{Hits@3} & \textbf{Hits@5} & \textbf{MRR@5} \\
\midrule
\multicolumn{15}{c}{\textit{Traditional Retrieval Models}} \\
BM25    & 49.13 & 42.42 & 11.42 & 20.07 & 72.72 & 81.13 & 62.42 & 58.43 & 66.67 & 9.36 & 13.92 & 92.19 & 96.09 & 79.10 \\
\midrule
\multicolumn{15}{c}{\textit{General Pre-trained Models}} \\
BERT    & 54.55 & 50.79 & 15.09 & 28.02 & 77.27 & 81.82 & 66.06 & 65.71 & 77.60 & 9.87  & 16.07 & 95.31 & 96.88 & 90.23 \\
RoBERTa & 55.85 & 53.33 & 15.67  & 28.34 & 77.56 & 82.91 & 65.45 & 66.84 & 80.23 & 10.12 & 16.13 & 95.47 & 97.28 & 90.55 \\
BGE     & 57.29 & 51.52 & 16.98 & 28.55 & 77.27 & 86.36 & 65.68 & 68.98 & 81.34 & 11.11 & 17.42 & 95.60 & 98.11 & 90.51 \\
\midrule
\multicolumn{15}{c}{\textit{Legal-Specific Pre-trained Models}} \\
Lawformer & 54.58 & 50.79 & 15.95 & 26.90 & 77.27 & 90.91 & 62.80 & 70.44 & 80.46 & 11.09 & 16.90 & 96.06 & 97.43 & 91.80 \\
SAILER    & 58.28 & 53.51 & 18.62 & 27.92 & 71.96 & 80.37 & 67.90 & 73.60 & 84.37 & 12.44 & 17.17 & 95.63 & 98.50 & 92.84 \\
\midrule
\multicolumn{15}{c}{\textit{Legal-Specific Re-ranking Models}} \\
KELLER    & \textbf{61.81} & 55.88 & 19.01 & 29.52 & 83.81 & 88.57 & 68.20 & 76.22 & 85.62 & 11.92 & 19.55 & 95.94 & 98.71 & 93.02 \\
\midrule
\textbf{Ours} & 58.61 & \textbf{56.06} & \textbf{26.13} & \textbf{33.88}$^\dagger$ & \textbf{95.45}$^\dagger$ & \textbf{95.45}$^\dagger$ & \textbf{71.97} & \textbf{76.58} & \textbf{86.58} & \textbf{12.73} & \textbf{19.62} & \textbf{97.48} & \textbf{99.37} & \textbf{93.52} \\
\bottomrule
\end{tabular}
}
\caption{Retrieval performance on LeCaRD and LeCaRDv2 datasets. The best results are highlighted in \textbf{bold}. $\dagger$ indicates statistically significant improvements over the strongest baseline (KELLER) with $p < 0.05$. Note that on the LeCaRD dataset, our method achieves significantly higher recall and hit rates, demonstrating superior robustness compared to KELLER despite a lower MAP.}
\label{tab:main_results}
\end{table*}

\subsection{Performance Comparison with Baselines (RQ1)} \label{sec:rq1_results}

Table~\ref{tab:main_results} presents the retrieval performance of our proposed framework compared to state-of-the-art baselines on the LeCaRD and LeCaRDv2 datasets. We categorize the baselines into two groups: (1) \textbf{General Semantic Retrieval Models}, including sparse retrieval (BM25) and dense retrieval models (BERT, RoBERTa, BGE); and (2) \textbf{Legal-Specific Pre-trained Models}, including Lawformer, SAILER, and the previous state-of-the-art method, KELLER.

From the results, we observe distinct performance patterns across the two datasets:

\paragraph{Consistent Superiority on LeCaRDv2.} On the LeCaRDv2 dataset, previous methods like KELLER have established a high performance baseline (MAP $>$ 76\%), suggesting a potential ceiling effect. Despite this saturation, our method achieves \textbf{state-of-the-art performance across all seven evaluation metrics}. While the numerical margins are narrower due to the high baseline (e.g., improving MAP from 76.22\% to \textbf{76.58\%} and Hits@5 from 98.71\% to \textbf{99.37\%}), the consistency of these improvements confirms that our generative paradigm successfully generalizes to diverse legal scenarios. By explicitly modeling the hierarchical structure of legal charges and elements, our framework effectively retrieves cases that possess consistent juridical logic, even when lexical overlap is limited.

\paragraph{Robustness and Safety on LeCaRD.} On the LeCaRD dataset, our method demonstrates exceptional robustness, particularly in recall-oriented metrics. Most notably, our \textbf{Hits@3 reaches 95.45\%}, significantly outperforming the strongest baseline KELLER (83.81\%) by a margin of \textbf{11.64\%} and SAILER (71.96\%) by \textbf{23.49\%}. Statistical tests confirm that these improvements in Hits@3, Hits@5, and R@5 are significant ($p < 0.05$).
Furthermore, our method achieves a remarkable gain in Recall@3 (\textbf{26.13\%} vs. KELLER's 19.01\%), demonstrating a superior ability to cover relevant precedents.
It is worth noting that while KELLER achieves a higher MAP (61.81\%) compared to ours (58.61\%), our method dominates in terms of finding the \textit{correct} cases (Hits) rather than just ranking them (MAP). In real-world legal practice, avoiding "zero-recall" failures (where no relevant case is found in the top results) is often prioritized over precise ranking permutations. Our framework effectively mitigates this risk, ensuring a "safer" retrieval experience.

\paragraph{Effectiveness of Legal Indicator Injection.} Comparing general dense retrievers (e.g., BGE) with our method reveals a clear performance gap. General models struggle to distinguish subtle legal nuances. They often retrieve cases with high semantic similarity yet erroneous legal characterization, while our framework incorporates generated charges and legal elements as hard constraints to address this limitation.This confirms that incorporating explicit legal knowledge via our Generative Legal Inference module effectively bridges the semantic gap that traditional embeddings fail to capture.

\subsection{Mechanism Analysis and Ablation (RQ2)}
\label{sec:ablation_mechanism}

To illustrate the effectiveness of our framework components and understand the underlying ranking logic, we conduct architectural ablation studies and employ SHAP (SHapley Additive exPlanations) for feature interpretability.

\paragraph{Architecture Validity.}
As shown in Table~\ref{tab:ablation_architecture}, removing the MLP scorer (\textbf{w/o MLP}) causes a drastic MAP drop (-15.2\%), indicating that the relationship between semantic correctness (e.g., Charge accuracy) and lexical matching (BM25) is highly non-linear. A simple rule-based sum fails to balance these distinct signals. 
Furthermore, bypassing the fine-tuned student model (\textbf{w/o GenIR}) leads to performance degradation (MAP 76.58\% $\to$ 74.78\%). We attribute this to the \textbf{Standardization of Legal Terminology}: while the teacher LLM is powerful, it suffers from hallucinations (e.g., generating synonymous but non-existent terms). In contrast, the student model, trained on the ``Silver Standard'' data, aligns colloquial queries with the \textbf{standardized legal vocabulary}, ensuring the generated indicators are strictly \textit{retrievable}.

\begin{figure}[t]
    \centering
    \begin{subfigure}[b]{0.95\linewidth}
        \centering
        \includegraphics[width=\linewidth]{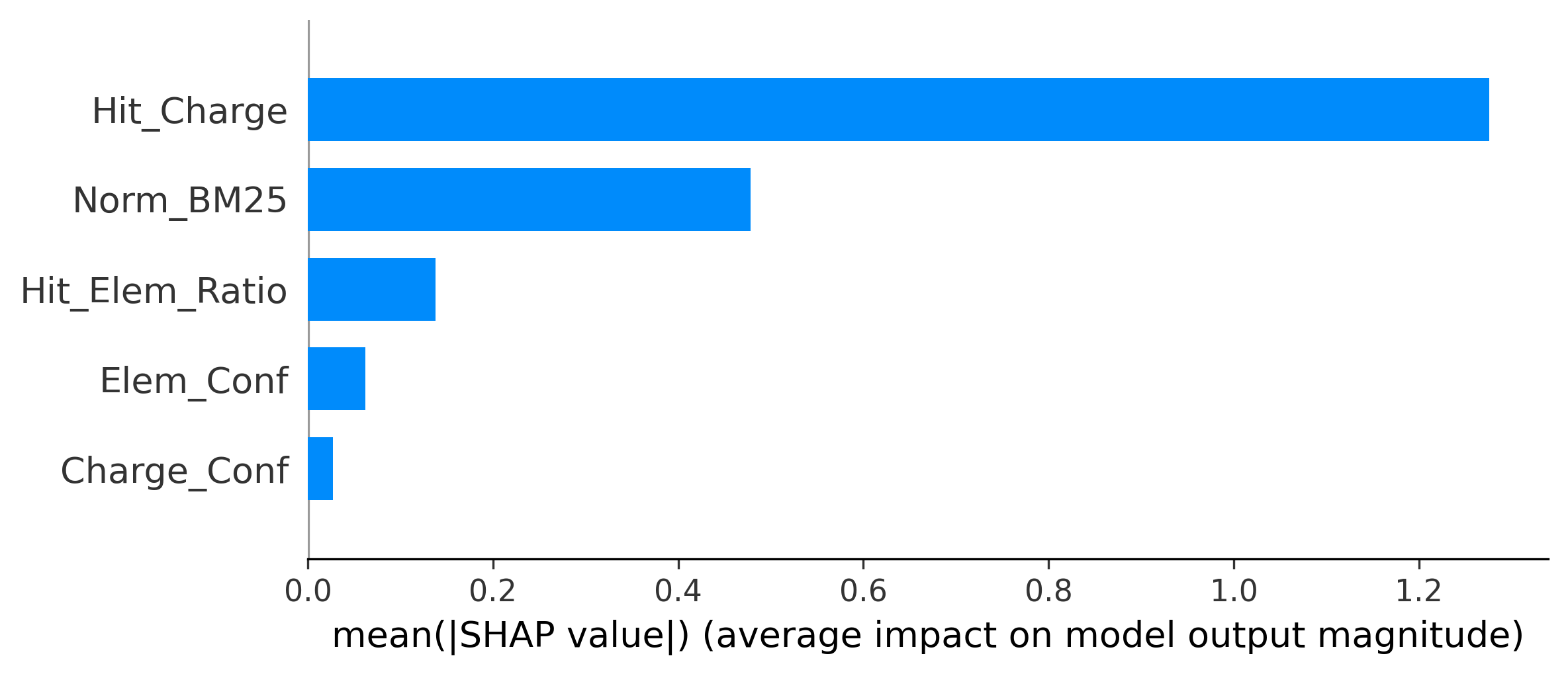}
        \caption{Global Feature Importance (Mean $|$SHAP$|$)}
        \label{fig:shap_bar}
    \end{subfigure}
    \par\bigskip 
    \begin{subfigure}[b]{0.95\linewidth}
        \centering
        \includegraphics[width=\linewidth]{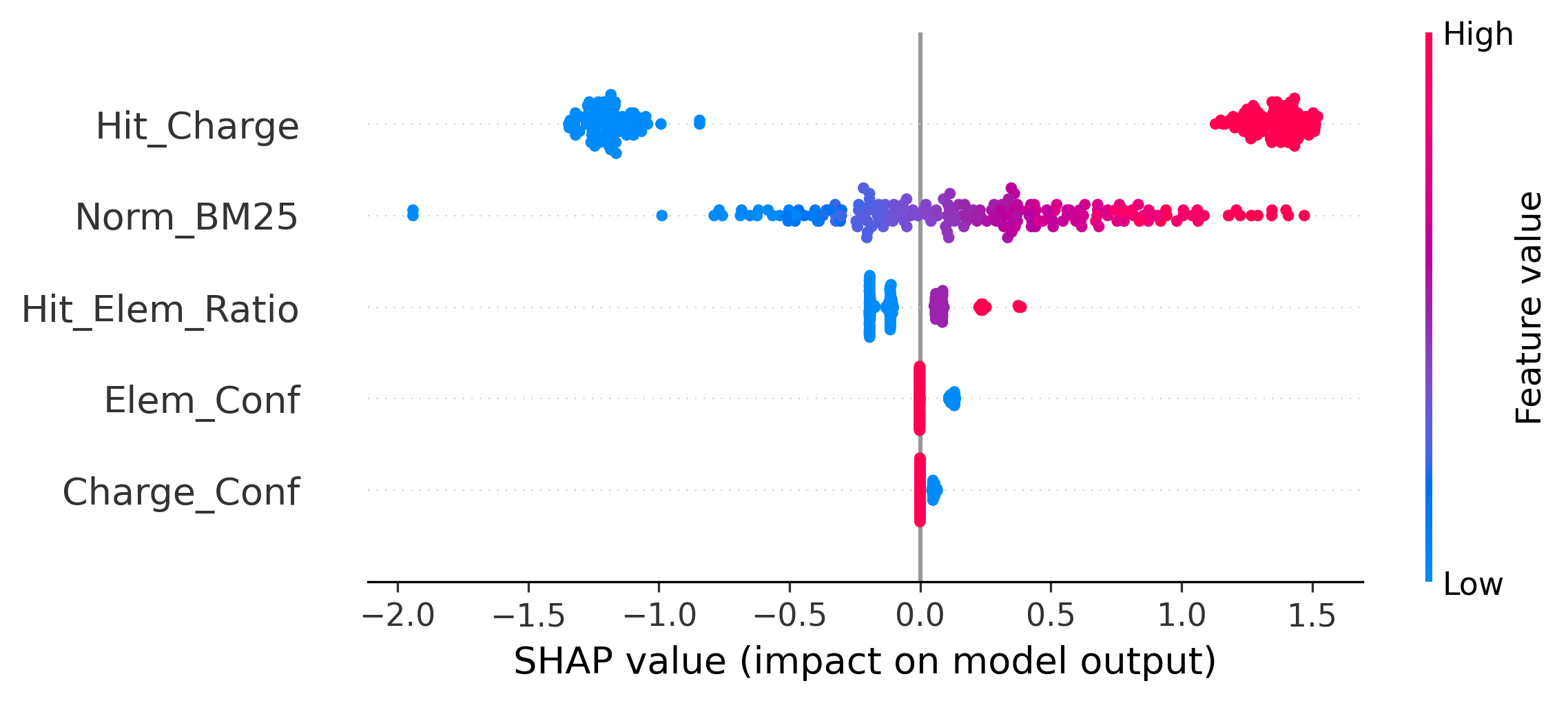}
        \caption{Detailed Feature Impact Distribution}
        \label{fig:shap_dot}
    \end{subfigure}
    \caption{\textbf{SHAP Interpretation of the MLP Scorer.} (a) shows \textit{Hit\_Charge} is the dominant factor. (b) reveals distinct roles: \textit{Hit\_Charge} acts as a decisive binary filter (clear separation), while \textit{Norm\_BM25} provides fine-grained calibration (continuous distribution).}
    \label{fig:shap_analysis}
    \vspace{-1.0em} 
\end{figure}

\paragraph{Interpretability of Ranking Features.}
To understand \textit{how} the MLP integrates these signals, we analyze the feature contributions in Figure~\ref{fig:shap_analysis}. 
\textbf{\textit{Hit\_Charge}} dominates the global importance, acting as a decisive "gatekeeper." As seen in Figure~\ref{fig:shap_dot}(b), a charge mismatch (blue dots) significantly penalizes the score, aligning with judicial logic: a case with the wrong charge is fundamentally irrelevant.
However, \textbf{\textit{Norm\_BM25}} ranks second, functioning as a fine-grained ranker to distinguish factually similar candidates within the same charge category.
This validates our \textbf{Complementary Ranking Strategy}: the model relies on generative signals for logical filtering and lexical matching for factual alignment.
Further detailed feature ablation (e.g., assessing the impact of removing Lexical features entirely) is provided in Appendix~\ref{sec:appendix_feature_ablation}.

\begin{table}[t]
\centering
\resizebox{\linewidth}{!}{%
\begin{tabular}{l|ccccccc}
\toprule
\textbf{Method} & \textbf{MAP} & \textbf{P@3} & \textbf{R@3} & \textbf{R@5} & \textbf{Hits@3} & \textbf{Hits@5} & \textbf{MRR@5} \\
\midrule
\textbf{Ours (Full Model)} & \textbf{76.58} & \textbf{86.58} & \textbf{12.73} & \textbf{19.62} & \textbf{96.86} & \textbf{99.37} & \textbf{93.52} \\
\midrule
\textit{Architecture Variants} & & & & & & & \\
\quad w/o GenIR (LLM+Prompt only) & 74.78 & 84.12 & 12.00 & 19.01 & 96.23 & 98.11 & 91.14 \\
\quad w/o MLP (Rule-based Rank) & 61.38 & 72.45 & 10.15 & 15.52 & 94.34 & 96.23 & 84.22 \\
\bottomrule
\end{tabular}%
}
\caption{Ablation study on model architecture. We compare the Full Model against variants without the finetuned generative module (using only LLM+Prompt) and without the MLP scorer (using rule-based ranking).}
\label{tab:ablation_architecture}
\end{table}

\subsection{Data Efficiency and Robustness (RQ3)}
\label{sec:rq3}

In practical legal scenarios, obtaining high-quality labeled data is often the bottleneck. To evaluate the robustness of our framework under data scarcity, we trained our model using stratified subsets of the training set, ranging from 10\% to 100\%.

\begin{figure}[t]
    \centering
    \includegraphics[width=\linewidth]{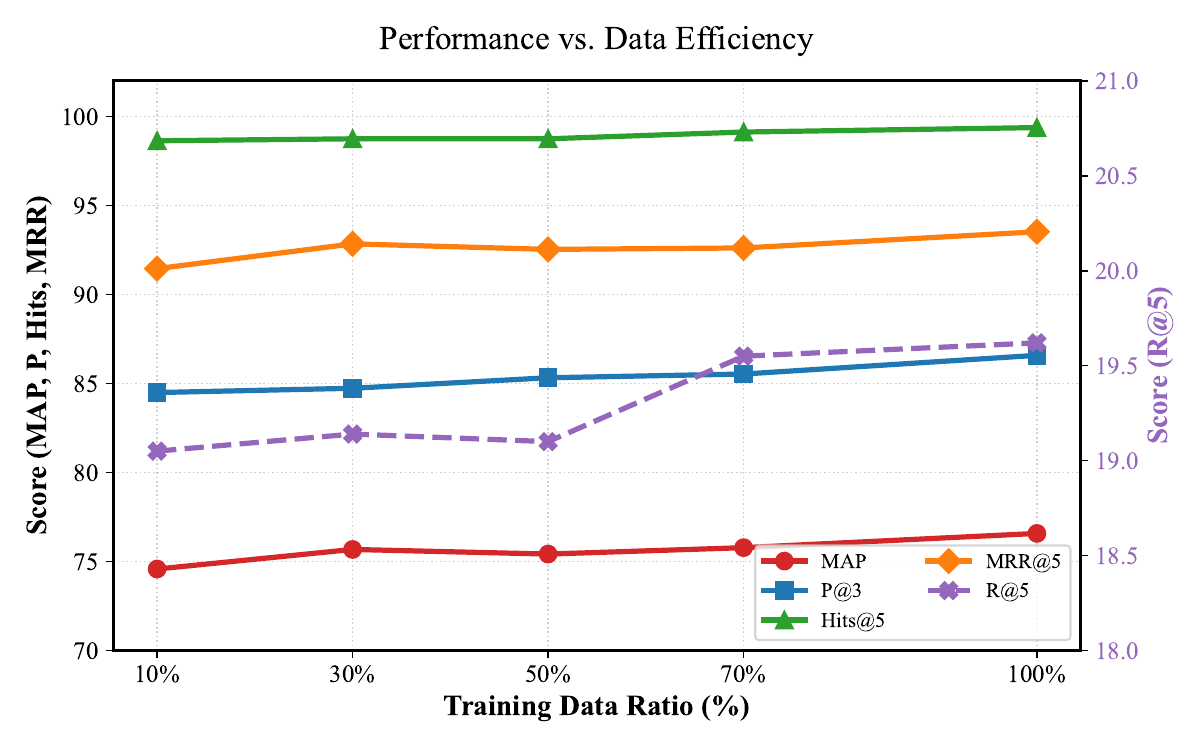} 
    \caption{Performance trends on LeCaRDv2 across varying training data ratios ($10\%\sim100\%$). The model demonstrates rapid convergence, achieving near-optimal performance (e.g., Hits@5 $>99\%$) with only $30\%$ of the data. Key metrics (left axis) remain stable, while Recall@5 (right axis) shows a slight continuous gain.}
    \label{fig:data_efficiency}
\end{figure}

\paragraph{Low-Resource Dominance.} 
As illustrated in Figure~\ref{fig:data_efficiency}, our method exhibits exceptional data efficiency. Notably, the performance curves for precision-oriented metrics (MAP, P@3, MRR@5) flatten rapidly, indicating saturation. Even with only \textbf{10\%} of the training data, our model achieves a MAP of 74.58\%, which already outperforms the full-data versions of strong baselines like SAILER (73.60\%) and Lawformer (70.44\%). The performance gain diminishes as data increases, reaching near-optimal results (75.68\% MAP) with just 30\% of the data.

\paragraph{Trend Analysis.} 
While ranking metrics stabilize early, Recall@5 (purple dashed line, right axis) shows a slight but continuous improvement as data increases. This suggests that while the model quickly learns the core logic for identifying top candidates, increased data scale further helps in covering long-tail relevant cases. Detailed numerical results are provided in Table~\ref{tab:data_efficiency} (~\ref{app:data_efficiency}).

\paragraph{Why Does mT5 Learn So Fast?} 
We attribute this rapid convergence to two primary factors:
\begin{itemize}
    \item \textbf{Scale of LeCaRDv2:} Although we use a small ratio (10\%), the absolute volume of the LeCaRDv2 dataset is large enough to provide sufficient supervision signals for finetuning the pre-trained mT5 backbone.
    \item \textbf{High Intra-Class Homogeneity:} Legal documents differ significantly from general open-domain texts. Cases sharing the same charge exhibit massive repetitions in legal phrasing and logical structures. A small subset of documents is sufficient for the model to capture the mapping rules between factual descriptions and charges, generalizing effectively without requiring extensive memorization of unique case details.
\end{itemize}

\subsection{Impact of Hierarchical Latent Factorization (RQ4)}
\label{sec:rq4}

To validate the necessity of hierarchical modeling, we compare our \textbf{Hierarchical Generation} against an \textbf{Independent Generation} baseline (where charges and elements are predicted separately). As shown in Table~\ref{tab:hierarchical_constraint}, the hierarchical strategy yields consistent improvements across all metrics (e.g., +1.87\% MAP). This improvement stems from two factors:
(1) \textbf{Chain-of-Logic}: In legal reasoning, the \textit{Charge} naturally restricts the scope of \textit{Constitutive Elements}. By modeling $P(\text{Elements} | \text{Query}, \text{Charge})$, the charge acts as a \textbf{semantic anchor}, filtering out irrelevant elements (e.g., violent details in property crimes).
(2) \textbf{Contextual Guidance}: Although error propagation is a potential risk, results show that the charge serves as a strong prior. It resolves ambiguities in vague queries and prevents hallucinations by enforcing top-down constraints, outweighing the impact of prediction errors.

\begin{table}[t]
\centering
\resizebox{\linewidth}{!}{%
\begin{tabular}{l|ccccccc}
\toprule
\textbf{Method} & \textbf{MAP} & \textbf{P@3} & \textbf{R@3} & \textbf{R@5} & \textbf{Hits@3} & \textbf{Hits@5} & \textbf{MRR@5} \\
\midrule
\textbf{Full Model (Hierarchical)} & \textbf{76.58} & \textbf{86.58} & \textbf{12.73} & \textbf{19.62} & \textbf{96.86} & \textbf{99.37} & \textbf{93.52} \\
Independent Generation & 74.71 & 83.23 & 11.73 & 18.44 & 94.97 & 97.48 & 92.53 \\
\bottomrule
\end{tabular}%
}
\caption{Comparison between Hierarchical (Two-Step) and Independent Generation strategies.} 
\label{tab:hierarchical_constraint}
\end{table}

\section{Conclusion}

In this paper, we presented \textbf{GLIER}, a novel framework designed to bridge the semantic gap in Legal Case Retrieval by mimicking the cognitive process of legal experts. By integrating an LLM-distilled generative inference module with a multi-view evidence fusion mechanism, our approach effectively aligns colloquial queries with professional legal structures. Extensive experiments on the LeCaRD and LeCaRDv2 benchmarks demonstrate that GLIER achieves excellent performance, consistently outperforming strong baseline models. Furthermore, our framework exhibits exceptional data efficiency, maintaining high retrieval quality even when trained with only 10\% of the available data. Through detailed feature analysis, we also verified that the model successfully leverages judicial logic to improve both ranking precision and interpretability Future work will explore the application of this generative paradigm to a broader range of complex legal scenarios.

\section*{Limitations}

Despite the strong performance of GLIER, several limitations remain to be addressed:

First, the student model is initialized with mT5-base, which has a \textbf{maximum sequence length} constraint (e.g., 512 tokens). While our framework utilizes distilled legal indicators to mitigate the noise of long documents . extremely lengthy or verbose user queries may still suffer from information loss due to truncation, potentially leading to incomplete legal element inference.

Second, our "Silver Standard" dataset relies on the \textbf{knowledge distillation from a specific LLM} (ChatGLM). Although human evaluation confirms high accuracy, the inherent biases or occasional hallucinations of the teacher model could still propagate to the student retriever. 

Finally, our experiments are primarily conducted on the LeCaRD series datasets, which are based on the \textbf{Chinese legal system}. The applicability of GLIER's hierarchical structure (Charge $\rightarrow$ Elements) to other jurisdictions, such as Common Law systems that rely more heavily on precedent-based reasoning than codified statutes, requires further empirical validation.

\section*{Acknowledgments}
This work was supported by the National Natural Science
Foundation of China (No. 62376178), and the Priority Academic Program Development of Jiangsu Higher Education Institutions.

\bibliography{custom}

\appendix


\section{Datasets}
\label{sec:Datasets}
\begin{itemize}
    \item \textbf{LeCaRD}: Derived from criminal rulings of the Supreme People’s Court of China, LeCaRD consists of 107 query cases and 10,700 candidate cases. To ensure a fair comparison consistent with baselines like KELLER and SAILER, we adopted a standardized evaluation protocol: the dataset was split into training and testing sets with a ratio of \textbf{0.8/0.2} using a fixed random seed of \textbf{42}. Retrieval performance is evaluated by ranking the documents within the candidate pool.

    \item \textbf{LeCaRDv2}: This dataset scales up the evaluation with 800 query cases and 55,192 candidate cases. It introduces a wider variety of criminal charges and more intricate legal scenarios, serving as a comprehensive benchmark for generalization capability.
\end{itemize}

\section{Implementation Details}
\label{app:implementation}

We implement our framework using PyTorch and HuggingFace Transformers. 
For the Knowledge Distillation phase, we employ ChatGLM as the teacher model. To construct the "Silver Standard" dataset, we designed a strict prompt that instructs the model to extract 4--6 key legal elements using professional terminology. Crucially, the prompt explicitly forbids the inclusion of sentencing details (e.g., imprisonment terms) to prevent target leakage.

The Generative Student Model is initialized with \texttt{mT5-base}. We set the maximum source and target sequence lengths to 512 and 128, respectively. The model was trained on a single NVIDIA Tesla V100 (32GB) GPU using the AdamW optimizer for approximately 72 hours. During inference, we utilize beam search with a beam width of 3 to generate the legal indicators.

The Discriminative Scorer is a 3-layer MLP (Input $\to$ 64 $\to$ 32 $\to$ 1) with ReLU activation and Dropout ($p=0.1$). It is trained using Binary Cross-Entropy loss with a batch size of 64 and a learning rate of 1e-4. To handle data imbalance, we employ a hard negative mining strategy with a negative-to-positive ratio of 3:1. Furthermore, we apply \textit{Per-Query Normalization} to the BM25 scores, ensuring that lexical features are comparable across different queries regardless of their candidate pool distributions.

\section{Prompt for Legal Element Extraction}
\label{app:prompt_details}

To ensure the quality of the "Silver Standard" dataset, we designed a rigorous prompt for the teacher LLM (ChatGLM). As shown in Table~\ref{tab:prompt}, the prompt includes specific constraints to standardize terminology and, crucially, to prevent the leakage of sentencing information (which would otherwise compromise the retrieval task).

\begin{table}[h]
    \centering
    \small
    \renewcommand{\arraystretch}{1.3}
    \begin{tabular}{|p{0.95\linewidth}|}
        \hline
        \textbf{System Role:} You are a senior legal text analysis expert. Please extract the core "legal elements" for the given charge from the criminal case content. \\
        \hline
        \textbf{Input Data:} \\
        \quad $\bullet$ \textbf{Convicted Charge:} \{charge\_str\} \\
        \quad $\bullet$ \textbf{Case Content:} \{truncated\_text\} \\
        \hline
        \textbf{Extraction Constraints:} \\
        1. \textbf{Task Goal:} Extract \textbf{4 to 6} key legal elements that support the conviction. \\
        2. \textbf{Terminology:} Use professional legal terminology (e.g., "violation of transportation regulations", "causing death") rather than colloquial descriptions. \\
        3. \textbf{Anti-Leakage (Critical):} \textbf{Strictly Prohibit} the inclusion of specific sentencing outcomes (e.g., "fixed-term imprisonment", "detention", "compensation amount") or explicit conviction statements. \\
        4. \textbf{Content:} Do not simply repeat the charge name; ensure there is no semantic redundancy between elements. \\
        \hline
        \textbf{Output Format:} \\
        Please directly return a standard JSON object: \\
        \texttt{\{"legal\_elements": ["Element 1", "Element 2", ...]\}} \\
        \hline
    \end{tabular}
    \caption{The instruction prompt used for knowledge distillation via ChatGLM (translated from the original Chinese).}
    \label{tab:prompt}
\end{table}

To ensure data quality, we conducted a human evaluation on 100 stratified samples. Two legal graduate students assessed the LLM-extracted labels, yielding a Charge Accuracy of 97.0\% and Element Precision of 82.0\%, with a Cohen’s Kappa of 0.71 (substantial agreement). These results confirm that the distilled "Silver Standard" data provides reliable supervision signals, while the student model's superior performance suggests it further mitigates the remaining noise.

We employed a robust LLM cascade strategy to construct the ``Silver Standard'' candidates for LeCaRD and LeCaRDv2, primarily using \texttt{chatglm-flash} for knowledge extraction. To address generation failures in complex cases, we utilized \texttt{deepseek-R1} as a fallback model to successfully process the remaining 205 documents. Subsequently, a rigorous cleaning pipeline was applied to remove approximately 290 error instances ($\sim$2.7\%), filtering out data with non-unique identifiers, ambiguous semantic descriptions across different charges, and inaccurate summarizations, thus ensuring high-quality supervision signals.

\section{Detailed Feature Ablation Analysis(RQ2)}
\label{sec:appendix_feature_ablation}

To understand the contribution of different input signals to the final ranking, we conduct a comprehensive feature ablation study. We categorize the five input dimensions of the MLP scorer into three groups: \textbf{Lexical Features} (BM25 score), \textbf{Charge Features} (Charge Confidence \& Hit), and \textbf{Element Features} (Element Confidence \& Hit Ratio). Table~\ref{tab:ablation_feature_detail} summarizes the results on LeCaRDv2.

\paragraph{The Role of Lexical Signals: Granularity and Factual Anchoring.} 
A striking observation is that using \textit{Only Lexical Feature} (i.e., standard BM25 ranking) achieves a MAP of 58.43\%, which is notably higher than using purely legal generative features (50.23\%). We attribute this to the differing \textbf{discriminative granularity} of the signals. Legal indicators (Charges and Elements) are inherently \textbf{categorical}: once the model identifies a specific charge (e.g., ``Theft''), all candidate cases belonging to this charge receive similarly high confidence scores. This results in a lack of ranking resolution, as the generative module cannot distinguish between distinct factual contexts within the same crime category. In contrast, lexical matching (BM25) captures specific factual details (e.g., names, locations, object values), providing the necessary granularity to rank cases. Therefore, the lexical signal serves as the indispensable foundation for recall, preventing the ``ranking ties'' that occur when relying solely on broad legal categories.

\paragraph{Generative Signals as Semantic Gatekeepers.} 
Despite the lower standalone performance of generative features, their removal leads to catastrophic degradation in the Full Model (MAP drops from 76.58\% to 50.23\% when removing BM25, and to 60.19\% when removing Charge features). This confirms that while generative signals may lack fine-grained ranking capability, they function as critical \textbf{Semantic Gatekeepers}. They impose strict juridical constraints, filtering out ``Hard Negatives''—cases that share high lexical overlap with the query but differ fundamentally in legal characterization (e.g., \textit{Theft} vs. \textit{Embezzlement}).

\paragraph{Hierarchical Importance: Charge vs. Elements.} 
Comparing the legal features, removing Charge Features (``w/o Charge'') causes a significantly larger performance drop (MAP -16.39\%) than removing Element Features (MAP -2.98\%). This validates the hierarchical nature of legal relevance modeled by our framework. The Charge acts as a coarse-grained primary filter; a mismatch here renders the case irrelevant regardless of other similarities. The Element features serve as a fine-grained secondary verifier, helping to distinguish cases with the same charge but different constitutive requirements, providing the final boost to reach state-of-the-art performance.

\paragraph{Synergy of Hybrid Scoring.} 
The most significant finding is the \textbf{super-additive effect} of combining signals. The Full Model (76.58\%) drastically outperforms both ``Only Lexical'' (58.43\%) and ``Only Legal Features'' (50.23\%). This indicates that our MLP scorer successfully learns a \textbf{Complementary Ranking Strategy}: it relies on BM25 to locate factually similar candidates (High Recall), while leveraging the generated Charge and Element signals to strictly enforce legal consistency (High Precision). This synergy validates our design of fusing explicit factual knowledge with latent generative reasoning.

\begin{table}[h]
\centering
\resizebox{0.9\linewidth}{!}{%
\begin{tabular}{l|ccc}
\toprule
\textbf{Method} & \textbf{MAP} & \textbf{MRR@5} & \textbf{NDCG@5} \\ 
\midrule
\textbf{Full Model} (All Features) & \textbf{76.58} & \textbf{93.52} & \textbf{84.64} \\
\midrule
\multicolumn{4}{l}{\textit{Impact of Feature Removal}} \\
w/o Lexical Feature (BM25) & 50.23 & 66.03 & 51.69 \\ 
w/o Charge Features & 60.19 & 84.62 & 72.22 \\ 
w/o Element Features & 73.60 & 91.53 & 84.12 \\ 
\midrule
\multicolumn{4}{l}{\textit{Performance of Single Feature Group}} \\
Only Lexical Feature (BM25) & 58.43 & 79.42 & 65.13 \\ 
Only Charge Features & 48.26 & 66.55 & 49.30 \\ 
Only Element Features & 40.88 & 63.97 & 47.31 \\ 
\bottomrule
\end{tabular}%
}
\caption{Detailed ablation study of different features in the MLP scorer on LeCaRDv2.}
\label{tab:ablation_feature_detail}
\end{table}

\section{Detailed Experimental Setup for RQ3}
\label{app:data_efficiency}

In Section~\ref{sec:rq3}, we evaluated the data efficiency of our model. Here, we describe the sampling strategy and provide the detailed performance metrics.

\subsection{Stratified Sampling Strategy}
To ensure the statistical validity of the low-resource subsets, we did not perform simple random sampling. Instead, we employed \textbf{Stratified Sampling} based on charge categories. 
Given the long-tail distribution of crimes in the LeCaRDv2 corpus, random sampling might completely exclude rare charges from the training set. Therefore, for every charge type existing in the training corpus, we randomly sampled exactly $p\%$ (e.g., 10\%, 30\%) of the corresponding cases. 
This strategy ensures that the data distribution of the subset remains consistent with the full dataset, preserving the diversity of legal scenarios even in extremely low-resource settings.

\subsection{Full Results on Data Efficiency}

Figure~\ref{fig:data_efficiency} in the main text illustrates the performance trends. For precise comparison, Table~\ref{tab:data_efficiency} details the exact retrieval performance metrics across all training data ratios.


\begin{table}[h]
    \centering
    \small
    \setlength{\tabcolsep}{4pt}
    \begin{tabular}{lccccc}
    \toprule
    \textbf{Ratio} & \textbf{MAP} & \textbf{P@3} & \textbf{R@5} & \textbf{Hits@5} & \textbf{MRR@5} \\
    \midrule
    10\%  & 74.58 & 84.49 & 19.05 & 98.63 & 91.45 \\
    30\%  & 75.68 & 84.73 & 19.14 & 98.74 & 92.84 \\
    50\%  & 75.42 & 85.32 & 19.10 & 98.74 & 92.53 \\
    70\%  & 75.78 & 85.53 & 19.55 & 99.12 & 92.61 \\
    100\% & \textbf{76.58} & \textbf{86.58} & \textbf{19.62} & \textbf{99.37} & \textbf{93.52} \\
    \bottomrule
    \end{tabular}
    \caption{Performance evaluation on the LeCaRDv2 dataset with varying training data proportions.}
    \label{tab:data_efficiency}
\end{table}

\section{Supplementary Experiment on Backbone Robustness}
While our framework utilizes distilled legal indicators to mitigate the noise of long documents, extremely lengthy or verbose user queries may still suffer from information loss due to truncation, potentially leading to incomplete legal element inference. To further validate that the sequence length and model capacity of mT5-base do not dominate the overall effectiveness of GLIER, we conducted supplementary experiments by replacing mT5-base with Qwen2.5-7B-Instruct under the exact same framework. Specifically, we applied QLoRA fine-tuning to the 7B model and expanded the input context window to 1024 tokens. As shown in Table \ref{tab:backbone_ablation}, while the adoption of a significantly larger backbone with a longer context window does yield slight performance improvements (e.g., +0.0020 in MAP and +0.0084 in P@3), the overall gains are strictly marginal.  This suggests that GLIER’s effectiveness primarily stems from its structured latent inference formulation, rather than backbone scale or context length. 


\begin{table}[htbp]
  \centering
  \small
  \begin{tabular}{lccc}
    \toprule
    Metric   & Qwen2.5-7B & mT5-base & Diff \\
    \midrule
    MAP      & 0.7678 & 0.7658 & +0.0020 \\
    P@3      & 0.8742 & 0.8658 & +0.0084 \\
    R@3      & 0.1249 & 0.1273 & -0.0024 \\
    R@5      & 0.1962 & 0.1962 & 0.0000  \\
    Hits@3   & 0.9811 & 0.9748 & +0.0063 \\
    Hits@5   & 0.9937 & 0.9937 & 0.0000  \\
    \bottomrule
  \end{tabular}
  \caption{Performance comparison between Qwen2.5-7B-Instruct and mT5-base on LeCaRDv2}
  \label{tab:backbone_ablation}
\end{table}

\end{document}